\documentstyle[preprint,aps,prb]{revtex}

\draft

\begin{document}
\title{{\bf {\bf \it Ab-initio} determination of the localized/delocalized
$f$-manifold in UPd$_2$Al$_3$}}
\author{L. Petit$^{1}$, A. Svane$^{1}$, W.M. Temmerman$^{2}$, Z. Szotek$^{2}$, 
and R. Tyer$^{2,3}$}
\address{$^{1}$ Institute of Physic and Astronomy, University of Aarhus, \\
DK-8000 Aarhus C, Denmark\\
$^{2}$ Daresbury Laboratory, Daresbury, Warrington WA4 4AD, UK\\
$^3$ Department of Physics and Astronomy, University of Sheffield, Sheffield, S3 7RH, UK}

\date{\today}
\maketitle

\begin{abstract}
The electronic structure of UPd$_2$Al$_3$ is described 
using the self-interaction corrected local-spin-density approximation to
density functional theory. The
groundstate is found      to be characterized by the coexistence of
localized ($f^2$) and delocalized U $f$ electrons, 
in agreement with experimental
evidence. We observe significant difference in electronic structure 
between UPd$_2$Al$_3$ and the previously studied UPt$_3$ compound.
Even though a trend towards localization exists in UPt$_3$,
the total energies and the density of states at the Fermi level
favor a groundstate with localized $f^1$, rather than $f^2$
U ions. 
\end{abstract}



Heavy fermion (HF) systems are some of the most fascinating condensed matter materials.\cite{stewart}
Their highly correlated electrons give rise to large values 
of the low temperature specific heat coefficient, $\gamma$, which in certain cases 
is larger than that of a normal metal by more than a factor of 1000. This is associated
with highly enhanced effective quasiparticle masses, $m^{*}$, of some 1000 times the free 
electron mass. Recently new insight has 
been gained into the nature of these heavy quasiparticles
in the magnetically ordered heavy fermions UPd$_2$Al$_3$ and UPt$_3$.\cite{dressel}
This work has established that magnetic order is the prerequisite to formation of these 
heavy quasiparticles. Specifically, it is asserted that coupling of localized and
delocalized 5f electrons lies at the origin of heavy quasiparticles in 
UPd$_2$Al$_3$ and UPt$_3$.\cite{dressel,zwicknagl}

In the standard description, the large specific heat
coefficients can be correlated to a large density of states (DOS) 
of heavy quasiparticles at the Fermi level, 
i.e., a narrow resonance at the Fermi level resulting from the 
interaction of the localized $f$ states with the
conduction electrons. In the weak-coupling Kondo limit the atomic moments of the
localized states are 
individually screened by the conduction electrons. With increasing hybridization,
coherence sets in, leading to the Fermi liquid
of weakly interacting, extremely heavy quasiparticles.
However, the heavy fermion state will not form, if
the interactions between unscreenend or incompletely screened 
local moments induce a magnetic phase
transition, with the onset of magnetic order.
This is the case in many rare earth and actinide metals such as e.g. UPd$_{3}$.
Contrary to this picture, the new 
observation  by Dressel~\cite{dressel} indicates that for UPd$_2$Al$_3$ and UPt$_{3}$ 
the heavy quasiparticles are {\it only} 
created by magnetic excitations in the magnetically
ordered state, within the manifold of localised and delocalised f states.
In the present
 paper these localised and delocalised states in UPd$_2$Al$_3$ are 
determined from {\it ab initio} self-interaction corrected (SIC) local spin density (LSD)
calculations and discussed in comparison with UPt$_3$.\cite{petit}

In UPd$_2$Al$_3$, the specific heat coefficient is 
$\gamma$=140 mJ mol$^{-1}$ K$^{-2}$,~\cite{geibel} i.e. less extreme than the value for the
prototypical heavy fermion compound CeAl$_3$ ($\gamma$=1600 mJ mol$^{-1}$ K$^{-2}$)
but still considerably larger than the
value for a normal metal such as Pd ($\gamma$=9.4 mJ mol$^{-1}$ K$^{-2}$). 
At T$_N$=14.3 K, UPd$_2$Al$_3$ undergoes an
antiferromagnetic phase transition, with a resulting moment 
of 0.85 $\mu_B$ per U-ion. At the critical
temperature T$_c$=2 K,
~\cite{geibel} a superconducting phase transition occurs, as 
indicated by a sharp drop in
electrical resistivity, and a large jump in the 
specific heat.~\cite{geibel,caspary} The pronounced magnetic moment is
retained below T$_c$.~\cite{krimmel}
From the specific heat measurements,~\cite{caspary} muon spin
rotation measurements,~\cite{feyerherm} and neutron scattering studies,~\cite{mason} it
emerges that superconductivity and anti-ferromagnetism coexist in UPd$_2$Al$_3$. 
It has been established that magnetic excitations between two $f$-electrons localized at 
each site, and the remaining $f$-electron are causing superconductivity.~\cite{sato}
Recently it has been suggested~\cite{dressel,zwicknagl} that the heavy
fermion superconductor UPt$_3$ can be similarly described in terms of a 
coupling between the localized $f^2$ configuration and the delocalized $f$'s.

Applying density functional theory~\cite{kohn} (DFT) to such HF compounds is 
not straightforward. While this theory is formally exact for an arbitrary many-electron system, 
in its most widely used approximations, 
like the local spin density or generalized gradient approximation (GGA),
it relies on the exchange and correlation effects of a homogeneous electron gas, which is a 
rather poor reference system for heavy fermion compounds.
The LSD scheme has been very 
successful in describing the cohesive properties of 
solid state systems with itinerant valence
electrons. In the 4$f$ and 5$f$ systems, however, the band formation competes with a strong 
tendency towards localization due to large on-site $f$-$f$ correlations that go beyond the LSD. 
Therefore, it is not obvious to what extent the results from bandstructure calculations remain 
valid.  
Nevertheless, owing to the fact that the 5$f$ electrons in the U compounds are 
considerably less localized than for example the 4$f$ electrons in the corresponding rare-earth 
compounds, LSD is known to provide useful
electronic structure in the normal phase of these compounds. For instance, the Fermi surface of 
UPt$_3$ is well reproduced by LSD,\cite{albers} and for UPd$_2$Al$_3$
it was shown~\cite{sandratskii,knopfle} that the experimental de Haas
van Alphen results, and the measured magnetic moment, 
could be reproduced by calculations treating all the U
$f$-electrons as itinerant, hybridizing with the conduction electrons. 
The latter work~\cite{knopfle} concluded, however, that in
order to explain the very anisotropic magnetic susceptibility of
UPd$_2$Al$_3$, one needed to consider the coexistence of two different kinds of 5$f$ 
electrons, i.e., localized and itinerant ones. 

The self-interaction-corrected local-spin-density approximation (SIC-LSD),~\cite{perdew}
constitutes a scheme capable of treating both localized and delocalized electrons on an equal 
footing, by including an explicit energy
contribution for an electron to localize (the self-interaction correction).~\cite{temmerman} 
This localization energy is the sum of the
self-Coulomb and self-exchange-correlation energies of the f-state. For an extended state
this energy correction vanishes, but for a localized U 5$f$-state the self-interaction 
correction is of 
the order of 30 mRy. In the SIC-LSD application to actinide systems,\cite{petit,leon}
a particular localized $f^n$ configuration is assumed on each actinide ion, 
while all other degrees of freedom are allowed to form bands,
in a self-consistency cycle. 
The localized states loose their band-formation energy, but gain
the self-interaction correction energy which is subtracted from the
LSD total energy functional.  
On balance, comparing different localized $f^n$ scenarios, the global
energy minimum establishes which configuration is the groundstate. 
In the SIC-LSD picture, the localized
electrons are no longer available as band-forming valence electrons, 
and accordingly the valency of the actinide ion may be defined as\cite{nature,valency}
$N_{val}$=$Z$-$N_{core}$-$N_{semi}$-$N_{SIC}$, where Z is the atomic number (92 for U), 
$N_{core}$ is the number
of atomic core electrons (78), $N_{semi}$ is the number of semicore electrons (8), 
and $N_{SIC}$ is the number of localized electrons.
Thus, for UPt$_3$ the SIC-LSD scheme has revealed a groundstate characterized by localized
$f^1$ electron on U (pentavalent configuration), which is intermediate between the itinerant ($f^0$)
behavior in URh$_3$ and the more localized $f^2$ U ions found in UPd$_3$ and UAu$_3$
(tetravalent configurations).\cite{petit} In the present work a similar study  of
UPd$_2$Al$_3$ is reported. Three configurations have been investigated, 
assuming respectively localized $f^2$, $f^1$ and
$f^0$ configurations on the U sites. 
 

The SIC-LSD formalism has been implemented~\cite{temmerman,svane} 
within the tight-binding linear muffin-tin orbitals method (LMTO),~\cite{andersen,varenna} 
with the atomic sphere approximation (ASA). All relativistic effects, including spin-orbit
interaction, have been incorporated in the Hamiltonian. 
For UPd$_2$Al$_3$ two separate energy panels have been used, one for the U 6$s$ and 6$p$ semi-core
states, and one for the valence electrons.
Apart from the 4$d$, 5$s$ and 5$p$ orbitals on the Pd atom, 
and the 3$s$, 3$p$, and 3$d$ orbitals on the Al atom, the
valence panel has comprised the 7$s$, 6$d$ and 5$f$ orbitals on the U atom with the 7$p$ degrees of
freedom treated by downfolding.\cite{varenna} 


In figure \ref{energy} we show the results of the total energy 
calculations as a function of volume for respectively the hexavalent ($f^0$), pentavalent 
($f^1$), and tetravalent ($f^2$) configurations, with the $f^0$ simply corresponding to the 
standard LSD approximation. The
tetravalent groundstate configuration is seen to give
the lowest total energy, in agreement with the U$^{4+}$ configuration predicted by experiment. 
The global minimum occurs at a volume of 681 au$^3$, 
in good agreement with the measured value, V$_{exp}$=700 au$^3$. The minima for the $f^0$ and
$f^1$ configurations are shifted slightly to lower volumes, as the partial pressure of
itinerant $f$-electrons is negative. The exact values of the minimum volumes depend somewhat
on the ratio of atomic spheres adopted in the ASA, but the relative ordering of the minima
is preserved, i.e., the finding of the ground state as $f^2$ is robust in this respect.  
The present calculations use the fixed ratio of atomic spheres
as $R_U:R_{Pd}:R_{Al}=3.4:2.4:2.5$.

With two of the $f$-electrons localized in the tetravalent configuration, the DOS of 
UPd$_2$Al$_3$ around the Fermi level
differs considerably from the 
corresponding DOS in the LSD configuration, as shown in         Fig. \ref{upd2al3}.
Thus, in the itinerant scenario, with all $f$-electrons delocalized (Fig. \ref{upd2al3}a), we obtain a relatively
large partial $f$-DOS (dotted line) at the Fermi level of about 102 states/Ry, as compared to the
$f^2$ configuration with only 20 states/Ry (Fig. \ref{upd2al3}b). In the latter configuration, most of the
$f$ weight has been transferred into the two localized states, and the remaining delocalized 
$f$-states
hybridize with the $s$, $p$ and $d$
levels on the neighbouring Al and Pd sites. This strong hybridization between $f$-electrons and conduction
electrons which is observed in both the tetravalent and delocalized~\cite{sandratskii} configurations, can
explain the fact that the measured specific heat coefficient is only about 140
mJ mol$^{-1}$ K$^{-2}$ in UPd$_2$Al$_3$, i.e., we are dealing with moderately heavy itinerant electrons at the
Fermi level. Despite this qualitative agreement, we must however not forget that by allowing an $f$-electron to
delocalize the corresponding strong on-site $f$-$f$ correlations are ignored, 
and that we therefore might
considerably overestimate the effect of $f$-conduction electron band-formation. The effect of these correlations
becomes clear when we compare the calculated and experimental values for the specific heat coefficient.  
Here, in the tetravalent configuration
we obtain  $\gamma$=12 mJ mol$^{-1}$ K$^{-2}$, i.e. too small by a factor of 10 with respect to
experiment. The same discrepancy between calculation and measurement has been observed in other heavy fermion
compounds,~\cite{steiner,albers} the reason being that although the correlations are actually not strong enough
to localize
the $f$-electrons, they lead to a considerable narrowing of the hybridized $fpd$-band, resulting in a
specific heat coefficient that is enhanced with respect to the LSD calculations by a factor
of up to
20-30.\cite{steiner,runge} Thus we do expect to find too small a value for $\gamma$, and the fact that LSD value
is in better agreement ($\gamma_{LSD}$=23 mJ mol$^{-1}$ K$^{-2}$) with experiment is a result of the 
increased $f$-weight at the Fermi level due to the delocalization of the $f$-electrons, rather than
an indication of a better description of the electronic structure in terms of narrow bands. 

There are now strong indications that, in UPd$_2$Al$_3$, the existence of two localized $f$-electrons is the
prerequisite for the appearance of superconductivity and heavy quasiparticles.
~\cite{dressel,sato,sato1,jourdan,thalmeier1} 
The argument is that the itinerant heavy
quasiparticles are strongly coupled to the localized $f$ electrons, and interact effectively with each
other by the intermediary of magnetic exitations. It has been suggested that the
superconductivity in UPt$_3$ might be of similar origin.\cite{dressel}
Optical conductivity measurements by these authors
show a pronounced pseudogap in the
low temperature response, in both UPd$_2$Al$_3$ and UPt$_3$, which they relate to
magnetic correlations between localized and delocalized $f$-electrons. 
Zwicknagl {\it et al.}~\cite{zwicknagl} 
have shown that treating two of the $f$-electrons in UPt$_3$ as localized in LDA band-structure 
calculations, may reproduce the observed de Haas-van Alpen frequencies, and 
that the interaction between itinerant and localized $f$-electrons may lead to
mass enhancements of a factor $\sim 10$.                          
There are however noticeable differences in the
low temperature properties of UPd$_2$Al$_3$ and UPt$_3$.\cite{heffner} Thus even though antiferromagnetism and
superconductivity coexist in UPt$_3$ below T$_c$=0.55 K, the magnetic moment is very small, only
0.02 $\mu_B$.
Also, unlike in UPd$_2$Al$_3$, in UPt$_3$ the magnetic moment is affected by the superconducting phase
transition.~\cite{aeppli}

The SIC-LSD calculations of UPt$_3$~\cite{petit} also find indications in favour of coexisting localized and
delocalized $f$-electrons. However in those calculations, it is the pentavalent $f^1$ configuration, rather than
the tetravalent $f^2$ configuration, which turns out to be energetically most favourable. 
Furthermore, in the
pentavalent configuration a narrow $f$-peak is pinned to the Fermi level, in agreement with 
the heavy fermion character of this compound,
whilst in the tetravalent configuration the 
UPt$_3$ DOS at the Fermi energy vanishes, as is shown in Fig. \ref{tetv}.
In fact, according to the present
SIC-LSD calculations, in the scenario
with two localized $f$-electrons, UPt$_3$ becomes very similar to UPd$_3$, which is 
then difficult to reconcile with
experimental evidence,~\cite{visser} since UPd$_3$ is neither a heavy fermion
compound nor a superconductor. 
In UPd$_2$Al$_3$, as can be seen from Fig. \ref{energy}, the energy difference between 
the $f^2$ and $f^1$ configuration is 
$E(f^2)-E(f^1)\sim-6$ mRy.
For UPt$_3$, UPd$_3$ and
UAu$_3$ the corresponding energy differences are $+5$ mRy, $-5$ mRy and $-9$ mRy,
respectively.\cite{petit}
Thus with respect to localization, UPd$_2$Al$_3$ is quite similar to UPd$_3$ (tetravalent configuration),
and more localized than UPt$_3$. Experimental evidence seems to suggest that UPt$_3$ is close to an
antiferromagnetic instability,~\cite{visser} and consequently the $f$-electrons are situated at the borderline of a
localization/delocalization transition. In our SIC-LSD calculations for UPt$_3$, this instability is reflected by
a groundstate configuration with one extra delocalized $f$-electron as compared to UPd$_2$Al$_3$. 
However, correlations of this extra delocalized electron are most likely not 
adequately accounted for by the LSD approximation, and would plausibly need 
to be described in a theory involving fluctuating 
$f^1$ and $f^2$ configurations, 
such as the dynamical coherent potential approximation.\cite{dyncpa}
In particular, the nearly vanishing magnetic moment observed for UPt$_3$
is not reproduced in the SIC-LSD approach.
The SIC-LSD results indicate however that it is 
energetically not favourable to statically localize two $f$ electrons.
Hence, the quasiparticle model of Zwicknagl {\it et al.}~\cite{zwicknagl} 
might possibly still be valid for     UPt$_3$, although it
 seems to be more straightforwardly
applicable to UPd$_2$Al$_3$.

In summary, SIC-LSD total energy and DOS calculations of 
UPd$_2$Al$_3$ have been performed. A ground state of localized U $f^2$ shells has been obtained 
where additionally strongly hybridized states of $f$ character appear at the Fermi level. 
The SIC-LSD {\it ab initio} calculations provide direct evidence for the picture of coexisting localized and
delocalized $f$-electrons which is inferred from experiments 
(such as \onlinecite{dressel}) to explain
heavy quasiparticles and superconductivity.
The calculations do not confirm a similar degree of localization for the $f$-electrons in UPt$_3$
and therefore magnetic excitations in UPd$_2$Al$_3$ and UPt$_3$ are most likely to be different. 



This work has been
partially funded by the Training and Mobility Network on `Electronic Structure 
Calculation of Materials Properties and Processes for Industry and Basic Sciences'
(contract:FMRX-CT98-0178) and the Research Training Network 'Psi-k f-electron'
(contract:HPRN-CT-2002-00295).

\begin{figure}[tbp]
\caption{ Total Energy (in Ry/formula unit) versus volume for  UPd$_2$Al$_3$. Three scenarios are
considered for the U-ion : f$^0$ (empty circles), f$^1$ (filled circles), and f$^2$ (stars). The arrow indicates
the value for the experimental volume.}
\label{energy}
\end{figure}

\begin{figure}[tbp]
\caption{ DOS of UPd$_2$Al$_3$, (a) in the hexavalent
configuration, (b) in the tetravalent configuration. The  fat solid, dotted, thin dolid, and 
dashed lines correspond, respectively, to the total DOS, the projected U $f$ DOS, 
the projected Pd $d$ DOS, 
and the projected Al $p$ DOS.
The energy (in Ry) is relative to the Fermi level (energy zero). 
 }
\label{upd2al3}
\end{figure}

\begin{figure}[tbp]
\caption{Total DOS of UPt$_3$ in the tetravalent configuration, which is energetically
unfavorable compared to the pentavalent configuration.
The energy (in Ry) is relative to the Fermi level (energy zero), which falls in the
hybridization induced gap.} 
\label{tetv}
\end{figure}


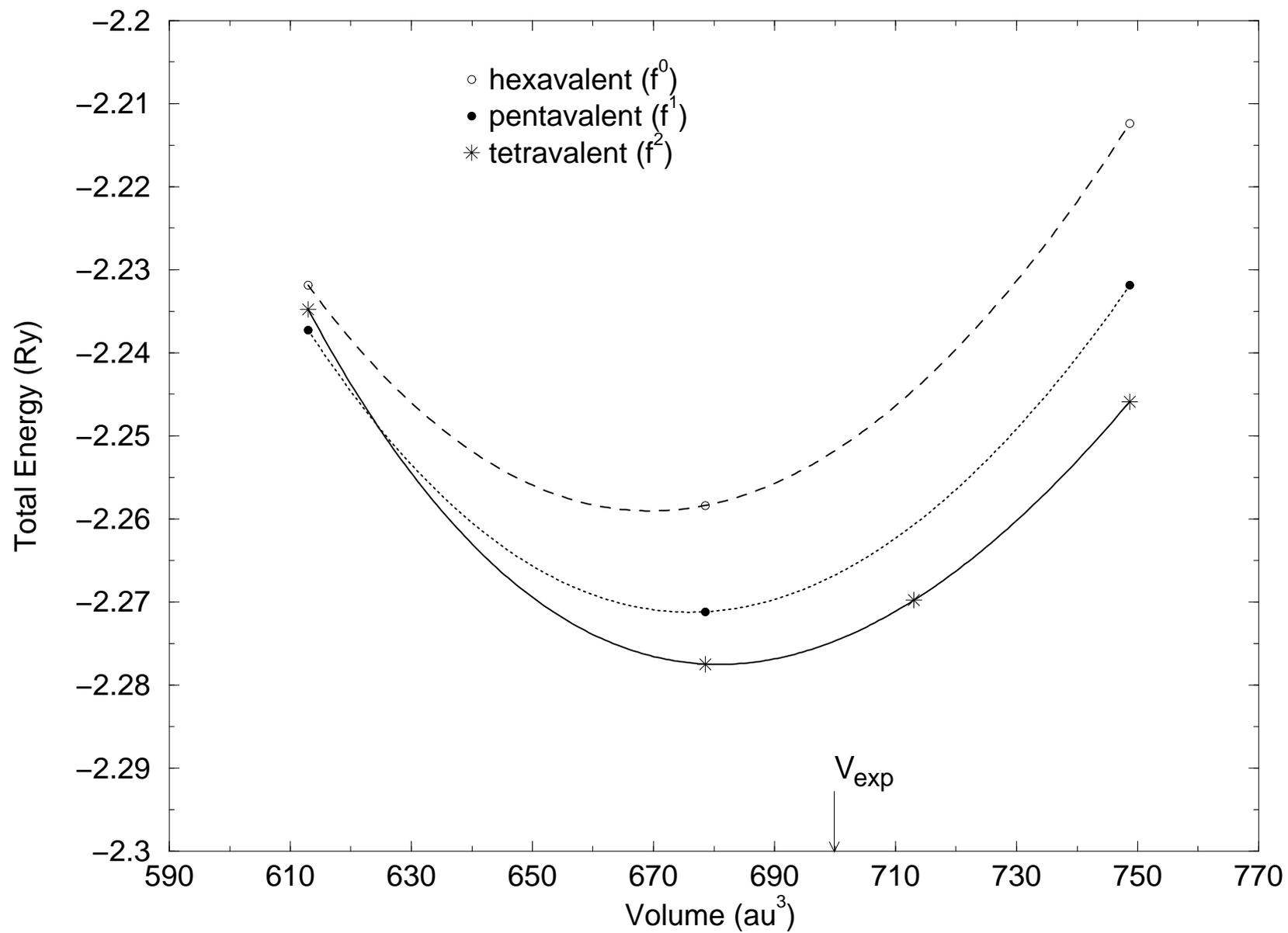
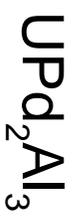
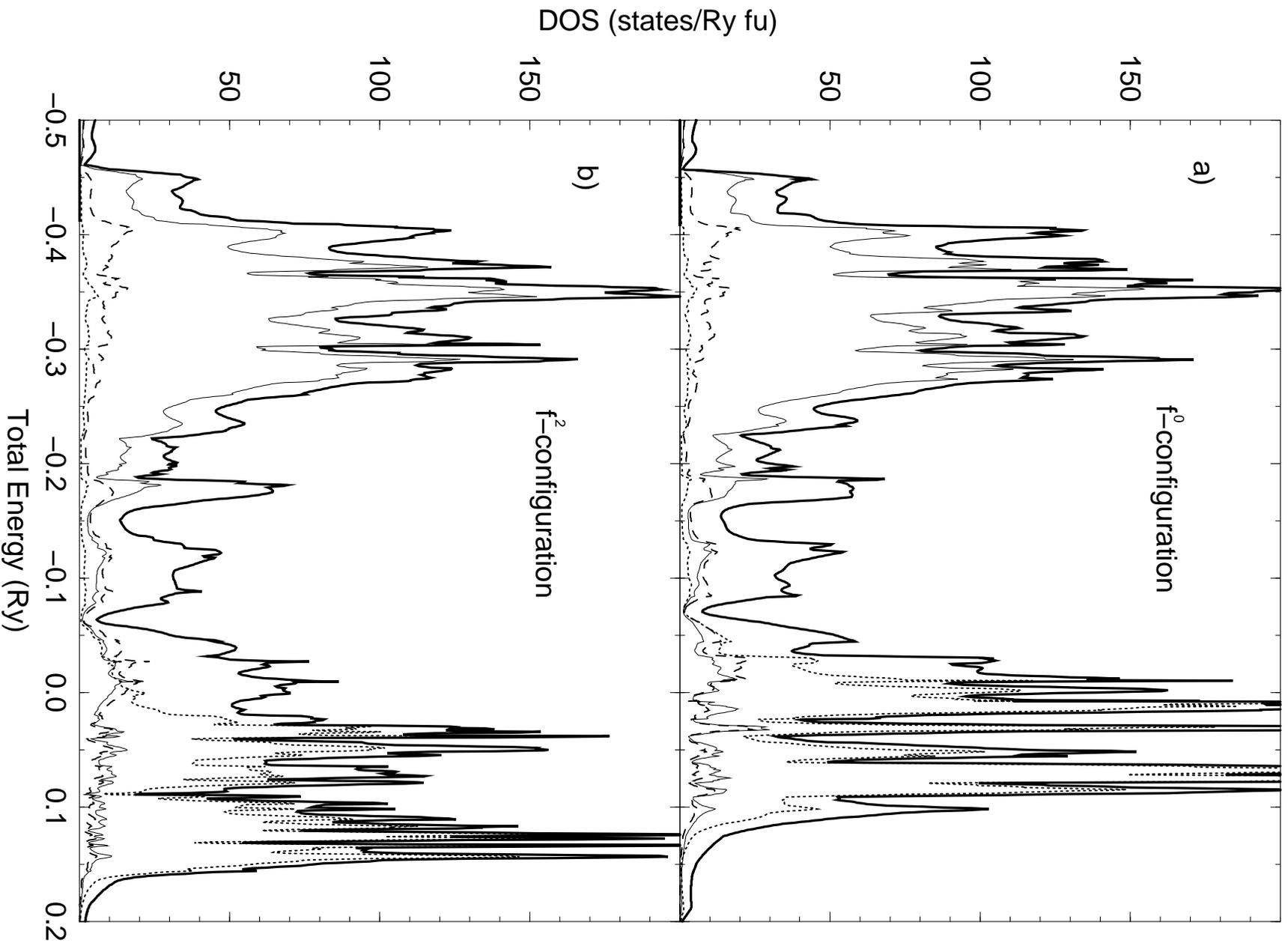
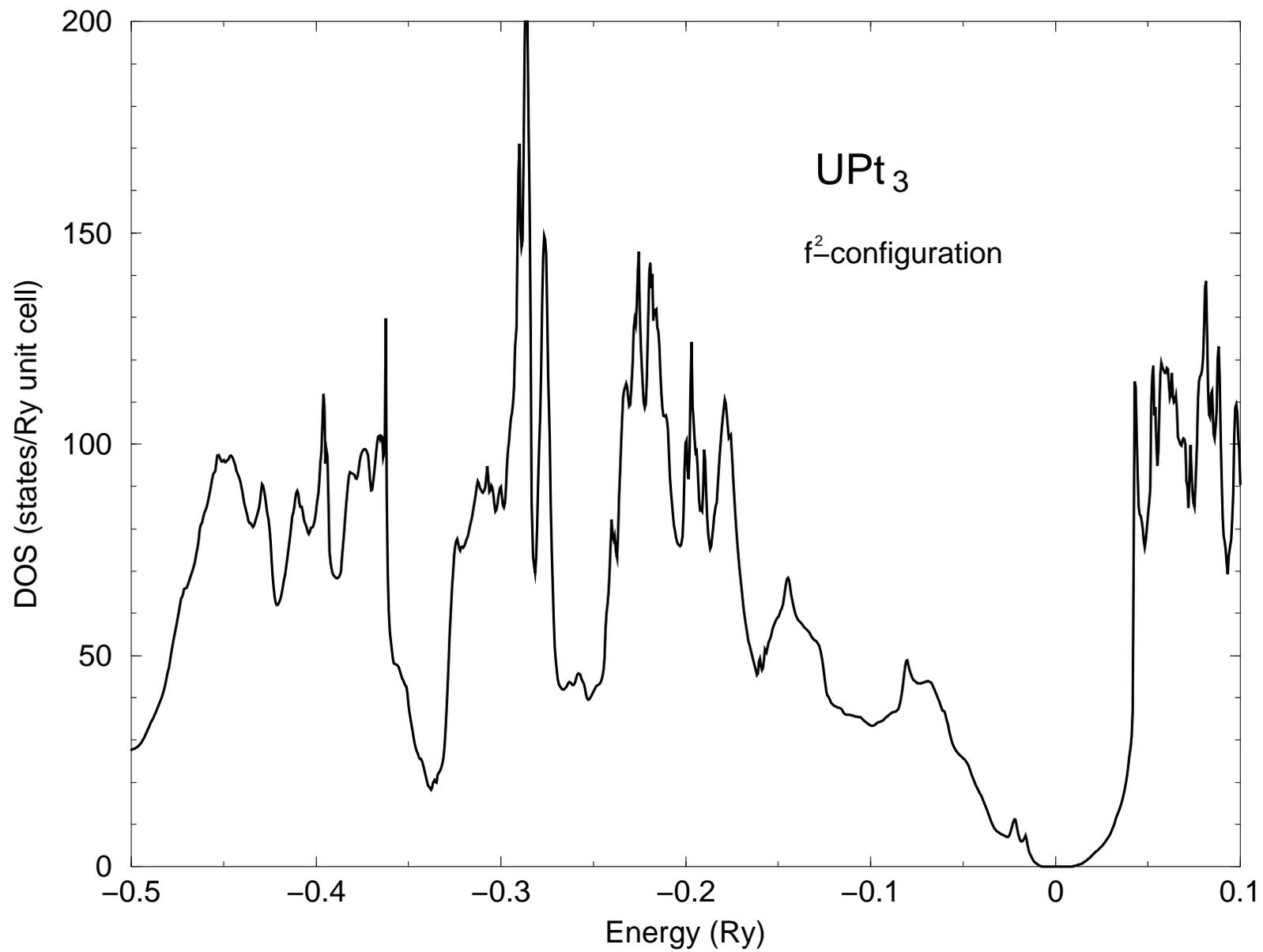
\end{document}